\documentclass[aps,superscriptaddress,showpacs,showkeys,a4paper,
superscriptaddress,10pt,nofootinbib,notitlepage,twocolumn]{revtex4-1}

\usepackage[utf8]{inputenc}
\usepackage[english]{babel}
\usepackage{amsmath}
\usepackage{amssymb}
\usepackage{amsfonts}
\usepackage{graphicx}
\usepackage{bm}
\usepackage{cancel}

\usepackage{newtxtext,newtxmath}
\usepackage[T1]{fontenc}

\usepackage[colorlinks,citecolor=blue,linkcolor=blue,urlcolor=blue]{hyperref}
\usepackage[tiny,center,uppercase]{titlesec}

\usepackage{tikz,xcolor}
\definecolor{lime}{HTML}{A6CE39}
\DeclareRobustCommand{\orcidicon}{%
    \begin{tikzpicture}
    \draw[lime, fill=lime] (0,0) 
    circle [radius=0.16] 
    node[white] {{\fontfamily{qag}\selectfont \tiny ID}};
    \draw[white, fill=white] (-0.0625,0.095) 
    circle [radius=0.007];
    \end{tikzpicture}
    \hspace{-2mm}
}

\newcommand{\orcidIDF}
{\href{https://orcid.org/0000-0003-0476-8634}{\orcidicon}}

\newcommand{\orcidODS}
{\href{https://orcid.org/0000-0002-2875-0140}{\orcidicon}}

\newcommand{\orcidShan}
{\href{https://orcid.org/0000-0003-2919-5414}{\orcidicon}}

\renewcommand{\vec}{\boldsymbol}

\newcommand{\thp}[2]{\vec #1\cdot\vec #2}
\newcommand\ri{\mathrm{i}}

\allowdisplaybreaks[2]

\begin{document}
	
\title{All-coupling solution for the continuous polaron problem in the
  Schr\"{o}dinger representation}

\author{I.\ D.\ Feranchuk\orcidIDF}
\email[Corresponding author: ]{iferanchuk@gmail.com}
\affiliation{Belaurusian State University, 4 Nezalezhnasty
  Ave., 220030, Minsk, Belarus}
\affiliation{Atomicus GmbH, Schoemperlen Str. 12a, 76185 Karlsruhe,
  Germany}

\author{N.\ Q.\ San\orcidShan}
\affiliation{Department of Physics, Faculty of Electricity and
  Electronics, Nha Trang University, Nha Trang, Vietnam}

\author{O.\ D.\ Skoromnik\orcidODS}
\email{olegskor@gmail.com}
\affiliation{Currently without university affiliation}

\begin{abstract}
  The solution for the large-radius Fr\"{o}hlich polaron in the
  Schr\"{o}dinger representation of the quantum theory is constructed
  in the entire range of variation of the coupling constant. The
  energy and the effective mass of the polaron are calculated by
  simple algebraic transformations and are analogous to the results
  found by Feynman on the basis of the variational principle for the
  path-integrals of this system. It allows us to solve the long-lived
  problem of the inequalities of the functional and operator
  approaches for the polaron problem. The developed method is
  important for other models of particle-field interaction including
  those ones for which the standard perturbation theory is divergent.
\end{abstract}

\maketitle

\section{Introduction}
\label{sec:introduction}
Presently it is well known that the polaron problem has broader
significance than simply a model of the interaction between an
electron and phonons in the ionic crystal as it was introduced by
Fr\"{o}hlich \cite{Froehlich1954}. It is important for description of
charge carriers in inorganic and organic matter interacting with ion
vibrations
\cite{doi:10.1080/00268976.2019.1567852,PhysRevB.98.045402}. The
corresponding electron-phonon interaction causes phase transitions,
including superconductivity and dominates the transport properties of
many metals and semiconductors (see for example, book
\cite{Devreese_book} and review \cite{Devreese_2009} and citations
therein).

Hamiltonian of the polaron problem is also important as a fundamental
model of the interaction between a particle and a quantum field. In
this problem various nonperturbative methods of quantum field theory
can be verified for the entire range of variation of the coupling
constant $\alpha$ of the interaction between an electron and a quantum
field \cite {Mitra198791}. Like any other quantum system the polaron
can be described both in the framework of the solution of the
Schr\"{o}dinger equation and by using the Feynman path-integral
formalism \cite{feynman1955slow}. The former approach allowed one to
introduce the idea of a self-localized polaron
\cite{pekar1963research} and to find the exact asymptotic value for
the ground state energy $E(\alpha)$ of the system in the strong
coupling limit $\alpha\gg1$ \cite{bogoliubov1950one}. While the latter
approach provided a uniform approximation for the energy of the system
in the whole range of the variation of the coupling constant
\cite{PhysRev.97.660}. It is important to notice that the solution for
the strong coupling ($\alpha\gg1$) is fundamentally different from the
solution in the case of weak coupling $\alpha\ll 1$ when the standard
perturbation theory can be applied \cite{Mitra198791}.

The great advantage of Feynman variational principle for the path
integrals is the possibility to calculate the polaron binding energy
$E(\alpha)$ as the continuous function for any $\alpha$. In addition
it allows one to find the lowest estimation for the polaron binding
energy in the intermediate coupling regime by the functional integrals
numerically. The effective diagrammatic quantum Monte Carlo algorithm
was developed for the Fr\"{o}hlich polaron in the path integral
representation \cite{mishchenko2007polarons, PhysRevB.62.6317,
  PhysRevB.97.134305}. It was considered as an important argument for
the advantage of the functional approach in the quantum field theory
in comparison with the Schr\"{o}dinger representation.

There were a lot of attempts \cite{Tokuda_1982, PhysRevB.27.6110,
  0022-3719-17-24-012, DASSARMA19851067, Lepine_1985} to calculate the
ground state energy with the help of variational principle for the
Schr\"{o}dinger representation of the polaron problem for all values
of the coupling constant $\alpha$ (all-coupling polaron). However, a
particular choice of the trial functions led to the singularity for
the energy of the system $E(\alpha)$ near the point $\alpha \simeq
7$. These results caused the discussion about existence of the “phase
transition” between two qualitatively different states of the polaron
(see review \cite{RevModPhys.63.63} on this problem). In a series of
papers cited in \cite{RevModPhys.63.63} it was proven that the
function $E(\alpha)$ is analytical for any value of $\alpha$ and the
``phase transition'' does not exist. Strict mathematical investigation
of the polaron problem in the strong coupling limit was recently
considered in the work \cite{lieb2020divergence}. However, it is
important to stress that till now no constructive computational
algorithm or trial wave function for variational approach are
developed for all-coupling solution of the polaron problem in the
Schr\"{o}dinger representation. The construction of such algorithm is
of great interest not only for the polaron problem but also for
non-perturbative description and analysis of the renormalization for
other models in the quantum field theory
\cite{Feranchuk2015}.

In the present paper we use operator method (OM) for calculation of
the ground state energy of the polaron problem for all values of the
coupling constant $\alpha$ in the Schr\"{o}dinger representation. The
OM was introduced in the paper \cite{feranchuk1982operator,
  Feranchuk1995370} and was effectively used later on for many quantum
systems \cite{Feranchuk2015, PhysRevA.96.052102}. It leads to the fast
convergent series for the solutions of the Schr\"{o}dinger
equation. This method was also applied for regular perturbation series
in the polaron problem \cite{feranchuk1982regular} but it was
considered only in the strong coupling limit.

In our work we for the first time demonstrate that in the case of
Fr\"{o}hlich Hamiltonian the two first terms of the OM series over
$\alpha$ lead to the function $E(\alpha)$ and the effective mass
$m_p(\alpha)$ of the polaron which fairly well coincide with Feynman’s
results. These functions can be calculated by rather simple analytical
expressions and lead to the correct asymptotic limits $\alpha \ll 1$
and $\alpha \gg 1$. In addition, good accuracy is achieved for
intermediate coupling with less numerical efforts as in comparison
with the path integral formalism. It seems to us that the results make
more clear and descriptive the question about the ground state of the
polaron and confirm the equivalence of the path integral and operator
approaches for description of quantum systems. Our analysis is
important for application of the self-localized states
for other models of the particle-field interactions even in the case
when conventional perturbation theory includes both the infrared and
ultraviolet divergences \cite{PhysRevD.92.125019}.

\section{Zeroth order approximation for the ground state
  energy} 
\label{sec:model_description}
Let us examine the Fr\"{o}hlich Hamiltonian for the system consisting
of a nonrelativistic electron that interacts with a quantum field of
optical phonons
\begin{align}\label{eq:1}
  \hat H &= \frac{\hat{\vec p}^2}{2}
           + 2^{5/4}\sqrt{ \frac{ \pi \alpha}{ \Omega}}
             \sum_{\vec k}\frac{\hat{q}_{\vec k}}{k} e^{\ri\thp{k}{r}}
           + \sum_{\vec k}\hat{c}^\dag_{\vec k} \hat{c}_{\vec k}.
\end{align}
Here the natural units with $\hbar = c = m = 1$ are chosen;
$\hat{c}^\dag_{\vec k} $ and $\hat{c}_{\vec k}$ are the phonon
creation and annihilation operators and $\hat{q}_{\vec k}$ is the
coordinate operator of the phonon field
\begin{equation*}
  \hat{q}_{\vec k} = \frac{1}{\sqrt{2}}(\hat{c}_{\vec k} + \hat{c}^{\dag}_{-\vec k}).
\end{equation*}

Let us also represent the electron coordinate $\hat{\vec r}$ and momentum
$\hat{\vec p}$ through the creation and annihilation operators, which allow
us later to perform all calculations in the algebraic form without
solutions of differential equations:
\begin{align}
  \hat{x}_{\lambda} =
  \frac{\hat{a}_{\lambda}+\hat{a}^{\dag}_{\lambda}}{\sqrt{2\omega}},\
  \hat{p}_{\lambda} = i \sqrt{\omega }
               \frac{\hat{a}^{\dag}_{\lambda} -  \hat{a}_{\lambda}}{\sqrt{2}},\
  [\hat{a}_{\lambda}, \hat{a}_{\mu}^{\dag}] = \delta_{\lambda \mu},\label{eq:2}
\end{align}
with a free parameter $\omega$. $\lambda$, $\mu$ numerate three
degrees of freedom of the particle. Recently it was also shown that
the polaron can be described in an algebraic form by $q$-deformed Lie
algebra \cite{PhysRevA.106.033321}.

As it was firstly shown by Ref.~\cite{pekar1963research}, the
electron-phonon interaction leads to the formation of the
self-localized state of the electron in the potential field of the
phonons. In order to take into account this effect we apply the
canonical transformation of the field operators
\begin{align}
\label{eq:3}
  \hat{q}_{\vec k} = u_{\vec k} + \hat{Q}_{\vec k};\
  \hat{c}_{ \vec k} = \frac{u_{\vec k}}{\sqrt{2}} + \hat{b}_{\vec k},
\end{align}
with the classical component of the field $u_{\vec k}$ , which will be
defined later.

The main idea of the OM is based on including in the zeroth-order
Hamiltonian $\hat H_0$ the terms from the full Hamiltonian that
commute with the operators of the number of the excitations
\begin{align}
  \hat n_{\lambda} &= \hat{a}^{\dag}_{\lambda}\hat{a}_{\lambda}, \label{eq:4}
  \\
  \hat N_{\vec k}  &= \hat{b}^{\dag}_{\vec k} \hat{b}_{\vec k}. \label{eq:5}
\end{align}
We now express $\hat H$ in terms of new operators. For this purpose we
use the operator identity
\begin{align}
  \exp
  \left[
  \frac{ik_{\lambda}(\hat{a}_{\lambda} + \hat{a}^{\dag}_{\lambda})}{\sqrt{2\omega}}
  \right]
  = e^{-\frac{k^2}{4\omega}}
  \exp\left(\frac{ik_{\lambda}  \hat{a}^{\dag}_{\lambda} }{\sqrt{2\omega}}\right)
  \exp\left(\frac{ik_{\lambda} \hat{a}_{\lambda}}{\sqrt{2\omega}}\right),
  \label{eq:6}
\end{align}
and split the Hamiltonian (\ref{eq:1}) into two parts
\begin{equation*}
  \hat H = \hat H_0 + \hat H_1,
\end{equation*}
where
\begin{align}
  \hat H_0
  &= \frac{3}{4}\omega +
  \frac{\omega}{2}(2\hat{a}^{\dag}_{\lambda}\hat{a}_{\lambda} - \hat{a}^{\dag}_{\lambda}
  \hat{a}^{\dag}_{\lambda} - \hat{a}_{\lambda}\hat{a}_{\lambda}) \nonumber
  \\
  &+ \frac{1}{2}\sum_{\vec k}\left[u_{\vec k}u_{-\vec k} +
    \frac{1}{\sqrt{2}}(u_{\vec k}\hat{b}^{\dag}_{\vec k} + u^*_{\vec k}\hat{b}_{\vec k})
     + \hat{b}^{\dag}_{\vec k}\hat{b}_{\vec k}\right] \label{eq:7}
  \\
  &+ \xi \sum_{\vec k}\frac{e^{-\frac{k^2}{4\omega}}}{k}
    \left[\hat{Q}_{\vec k} + u_{\vec k}
    \left(1 - \frac{k_{\lambda} k_{\mu}}{4\omega}
    (2\hat{a}^{\dag}_{\lambda}\hat{a}_{\mu} + \hat{a}^{\dag}_{\lambda}\hat{a}^{\dag}_{\mu}
    + \hat{a}_{\lambda}\hat{a}_{\mu})
    \right)\right],
    \nonumber
\end{align}
and
\begin{align}
  \hat H_1
  &= \xi\sum_{\vec k}\frac{e^{-\frac{k^2}{4\omega}}}{k}
  \Bigg[(\hat Q_{\vec k} + u_{\vec k})
    \left(\exp\left(\frac{ik_{\lambda}  \hat{a}^{\dag}_{\lambda} }{\sqrt{2\omega}}\right)
    \exp\left(\frac{ik_{\lambda} \hat{a}_{\lambda}}{\sqrt{2\omega}}\right)
        - 1 \right)
  \nonumber
  \\
  &+ u_{\vec k} \frac{k_{\lambda} k_{\mu}}{4 \omega}
    (2\hat{a}^{\dag}_{\lambda}\hat{a}_{\mu} + \hat{a}^{\dag}_{\lambda}\hat{a}^{\dag}_{\mu}
    + \hat{a}_{\lambda}\hat{a}_{\mu})\Bigg],
    \label{eq:8}
\end{align}
with
\begin{align*}
  \hat Q_{\vec k} = \frac{\hat{b}^{\dag}_{-\vec k}+ \hat{b}_{\vec k}}{\sqrt{2}};
    \quad \xi = 2^{5/4}\sqrt{\frac{ \pi \alpha}{ \Omega}}.
\end{align*}

The operator (\ref{eq:7}) is reduced to the diagonal form if we choose
the following values for the parameters $u_{k}$ and $\omega$
\begin{align}
  \omega
  &= \frac{4\alpha^2}{9 \pi}; \label{eq:9}
  \\
  u_k
  &= - 2^{5/4}\sqrt{ \frac{ \pi \alpha}{ \Omega}}\frac{e^{-\frac{k^2}{4\omega}}}{k};
  \label{eq:10}
\end{align}
and looks
\begin{align}
  \hat H_0 = - \frac{\alpha^2}{3 \pi} + \frac{4\alpha^2}{9 \pi}
  \hat{a}^{\dag}_{\lambda}\hat{a}_{\lambda}
           + \sum_{\vec k}\hat{b}^{\dag}_{\vec k}\hat{b}_{\vec k}. \label{eq:11}
\end{align}

Consequently the polaron ground state vector and energy in the zeroth
approximation are defined as follows
\begin{align}
  E_0^{(0)}
  &= - \frac{\alpha^2}{3 \pi} \label{eq:12}
  \\
  \hat{a}_{\lambda}|\psi_0 \rangle
  &= \hat{b}_{\vec k}|\psi_0 \rangle = 0. \label{eq:13}
\end{align}

The zeroth-order approximation alone does not provide the correct
asymptotic behavior for the energy of the system for the case of weak
coupling ($E\sim \alpha$). Therefore, we should take into account the
second-order correction, where we expect the restoration of the
correct asymptotic. We also notice here that this is a peculiar
property of the operator method where the second-order correction
restores the correct asymptotic behavior \cite{Feranchuk2015}.

\section{Second order approximation for the ground state energy}
\begin{figure*}
  \includegraphics[width=\columnwidth]{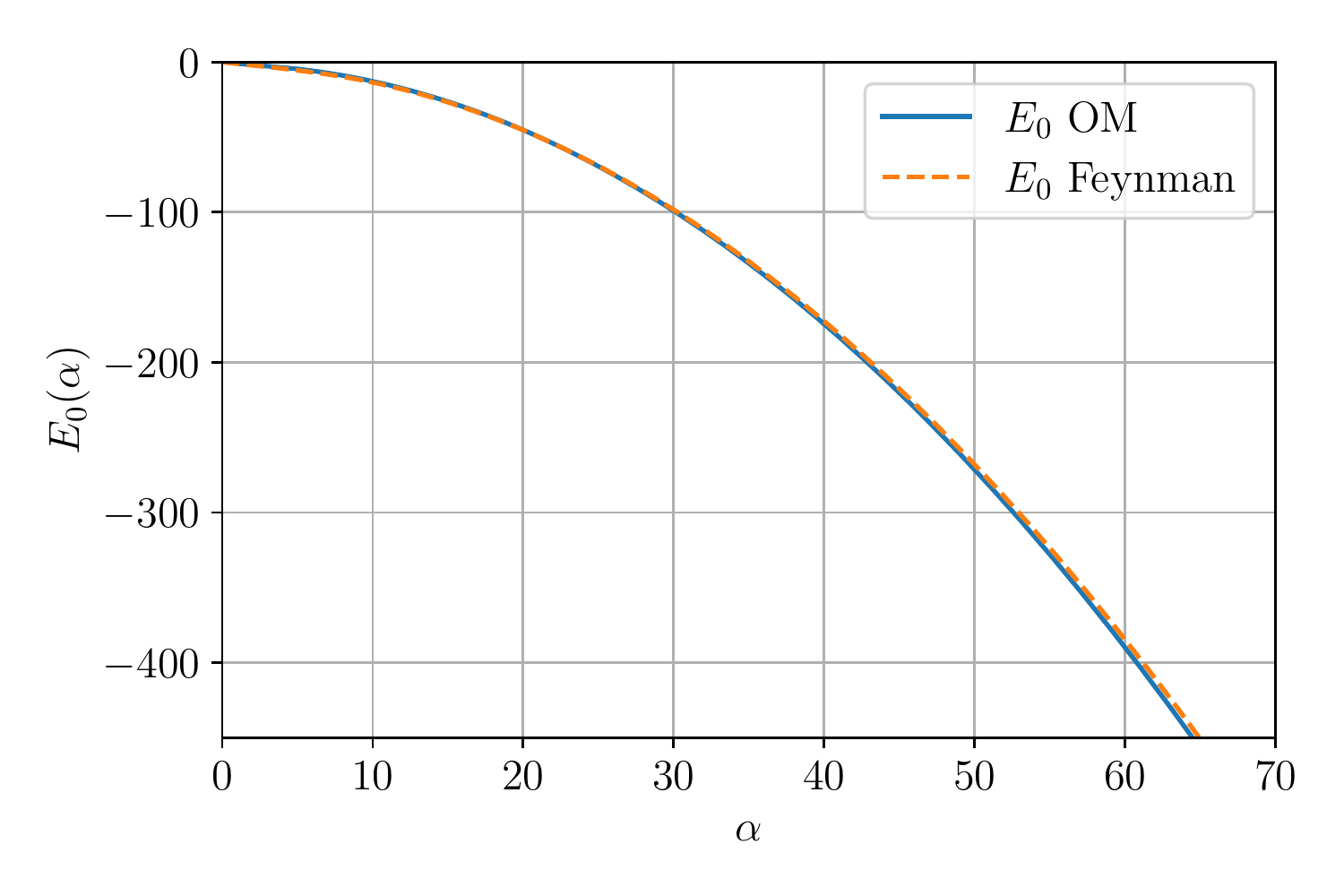}
  \includegraphics[width=\columnwidth]{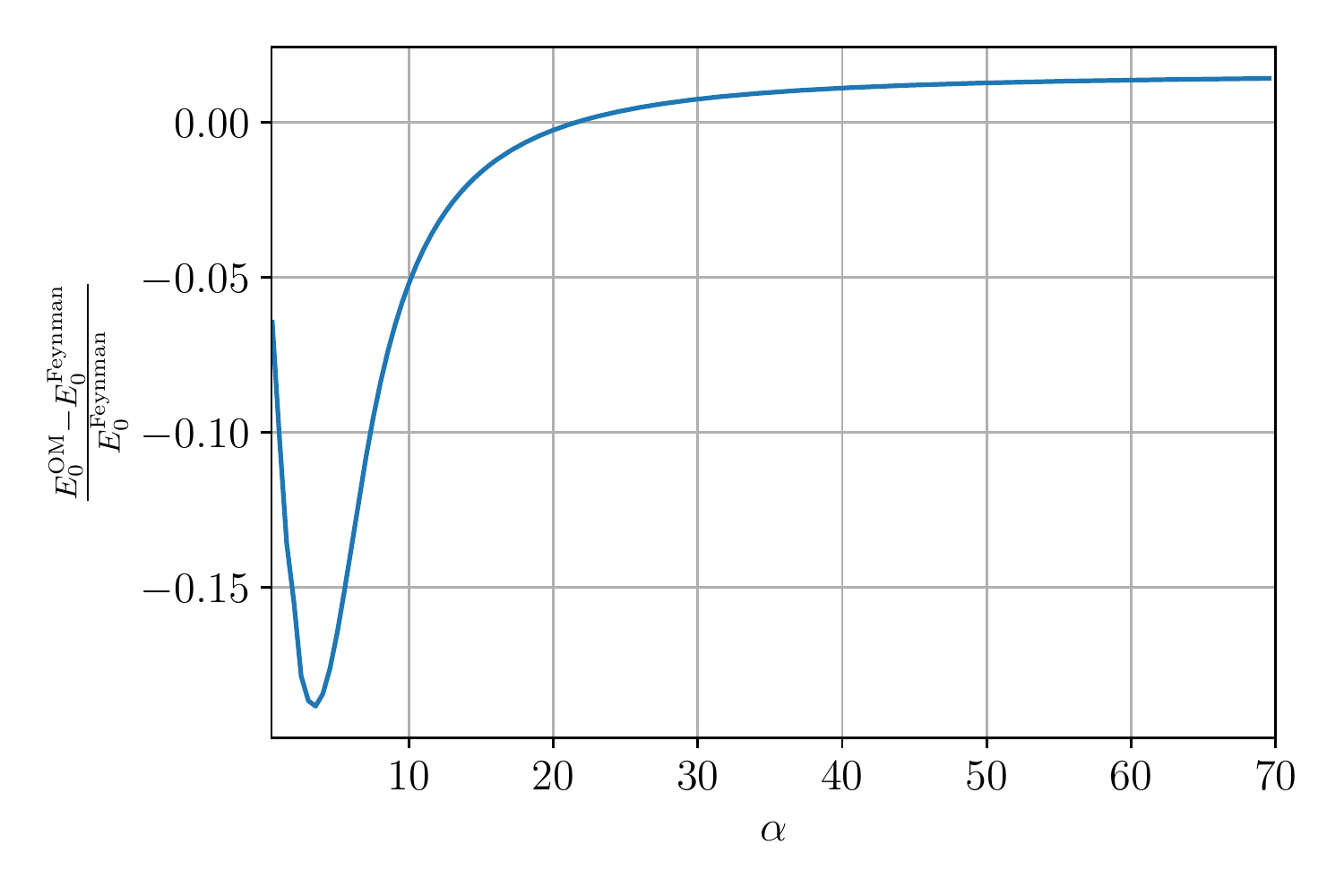}
  \caption{The polaron ground state energy and the relative
    difference as a function of $\alpha$ calculated by Feynman and our
    analytical formula}
  \label{fig:1}
\end{figure*}

Let us consider the perturbation series on the operator $\hat H_1$ for
the ground state energy. The first-order correction is equal to zero
identically and the second-order one is defined by the formula

\begin{align}
  \label{eq:14}
  E_0^{(2)}  = - \langle\psi_0|\hat H_1 [E_0^{(0)} -  \hat H_0]^{-1} \hat H_1  |\psi_0\rangle.
\end{align}
It is evident that the ground state should be excluded from the
resolvent spectrum. The calculation of (\ref{eq:14}) may be fulfilled in
the operator form if we use the integral representation
\begin{align}
  \label{eq:15}
  E_0^{(2)}  =  \int_0^{\infty} dx \langle\psi_0|\hat H_1
  e^{- (\omega \hat n + \sum_{\vec k} \hat N_{\vec k})x} \hat H_1  |\psi_0\rangle.
\end{align}

Let us calculate this value with the operator (\ref{eq:8}) represented
in the normal form
\begin{align}
  \hat H_1  |\psi_0\rangle = \xi \sum_{\vec k}\frac{e^{-\frac{k^2}{4\omega}}}{k}
  \Bigg[\left(   \frac{\hat{b}^{\dag}_{-\vec k}}{\sqrt{2}} + u_{\vec k}\right)
  \left(\exp\left(\frac{ik_{\lambda} \hat{a}^{\dag}_{\lambda} }{\sqrt{2\omega}}\right) - 1\right)
  \nonumber
  \\
  +  u_{\vec k}   \frac{k_{\lambda} k_{\mu}}{4 \omega}
  \hat{a}^{\dag}_{\lambda}\hat{a}^{\dag}_{\mu} \Bigg]|\psi_0\rangle,
  \label{eq:16}
\end{align}
\begin{widetext}
  \begin{align}
    [ e^{- (\omega \hat n + \sum_{\vec k} \hat N_{\vec k})x}  ]  \hat H_1  |\psi_0\rangle =
    \xi \sum_{\vec k}\frac{e^{-\frac{k^2}{4\omega}}}{k}
    \Bigg[\left(\frac{\hat{b}^{\dag}_{-\vec k}e^{-x}}{\sqrt{2}} + u_{\vec k}\right)
    \left(\exp\left(\frac{ik_{\lambda}  \hat{a}^{\dag}_{\lambda}e^{-\omega x} }{\sqrt{2\omega}}\right)
    - 1\right)
    +  u_{\vec k} \frac{k_{\lambda} k_{\mu}}{4 \omega}
    \hat{a}^{\dag}_{\lambda}  \hat{a}^{\dag}_{\mu}e^{-2\omega x} \Bigg]
    |\psi_0\rangle, \label{eq:17}
  \end{align}
  \begin{align}
    \langle\psi_0|\hat H_1
    = \xi \langle\psi_0|\sum_{\vec k_1}\frac{e^{-\frac{k_1^2}{4\omega}}}{k_1}
    \Bigg[\left(\frac{\hat{b}_{\vec k_1}}{\sqrt{2}} + u_{\vec k_1}\right)
    \left(\exp\left(\frac{ik_{1\lambda}  \hat{a}_{\lambda} }{\sqrt{2\omega}}\right)
    - 1\right)
    +  u_{\vec k_1}   \frac{k_{1\lambda} k_{1\mu}}{4 \omega}\hat{a}_{\lambda}  \hat{a}_{\mu} \Bigg],
    \label{eq:18}
  \end{align}
  \begin{align}
    \langle\psi_0|\hat H_1 e^{- (\omega \hat n + \sum_{\vec k} \hat N_{\vec k})x} \hat H_1  |\psi_0\rangle
    &= \xi^2 \Bigg[\sum_{\vec k}\frac{e^{-\frac{k^2}{2\omega}}}{k^2}  \frac{ e^{-x}}{2}
      \left( e^{\frac{k^2e^{-\omega x}}{2\omega}} - 1\right) \nonumber
    \\
    & \mspace{60mu}+ \sum_{\vec k}\sum_{\vec k_1}e^{-\frac{k^2+k_1^2}{4\omega}}
      \frac{u_k u_{k_1}}{k k_1}
      \left\{[e^{-( \frac{\vec k \vec k_1 }{2\omega}e^{-\omega x})} - 1]
      - \frac{(\vec k_{1}\vec k)^2}{4 \omega^2}e^{-2\omega x}\right\}\Bigg].
      \label{eq:19}
  \end{align}
\end{widetext}

We now calculate integrals over $\vec k$
\begin{align}
  \xi^2
  &\sum_{\vec k}\frac{e^{-\frac{k^2}{2\omega}}}{k^2}
    \frac{ e^{-x}}{2} ( e^{\frac{k^2e^{-\omega x}}{2\omega}} - 1) \nonumber
  \\
  &= 2^{5/2}  \frac{ \pi \alpha}{8\pi^3}4\pi\frac{ e^{-x}}{2}
    \int_0^{\infty}dk(e^{-\frac{k^2(1 - e^{-\omega x})}{2\omega}} - e^{\frac{- k^2}{2\omega}})  \nonumber
  \\
  &= 2^{5/2}\frac{\alpha e^{-x}}{4\pi}\frac{\sqrt{2\pi \omega}}{2}
    \left(\frac{1}{\sqrt{1- e^{-\omega x}}} - 1\right) \nonumber
  \\
  &= \alpha \sqrt{\frac{\omega}{\pi}}\left(\frac{1}{\sqrt{1- e^{-\omega x}}}- 1\right)e^{-x},
    \label{eq:20}
\end{align}
and
\begin{align}
  \xi^2
  &\sum_{\vec k}\sum_{\vec k_1}e^{-\frac{k^2+k_1^2}{4\omega}}
  \frac{u_k u_{k_1}}{k k_{1}}
  \left[\exp^{-( \frac{\vec k \vec k_1 }{2\omega}e^{-\omega x})} - 1\right]  \nonumber
  \\
  &= 2^5(\frac{    \alpha}{ 8\pi^2})^2\int\frac{d\vec k d\vec k_1}{k^2k_1^2}
    e^{-\frac{k^2+k_1^2}{2\omega}}
    \left(\exp^{-( \frac{\vec k \vec k_1 }{2\omega}e^{-\omega x})} - 1\right) \nonumber
  \\
  &=\frac{\alpha^2}{2\pi^4}(4\pi)^2 2\omega \int_0^{\infty} \int_0^{\infty}dXdY
    e^{- (X^2+ Y^2)}\left(\frac{\sinh XYt}{XYt} - 1\right) \nonumber
  \\
  &=\frac{\alpha^2}{2\pi^4}(4\pi)^2 2\omega\frac{\pi}{4}
    \left(\frac{2\arcsin t/2}{t} - 1\right) \nonumber
  \\
  &= \frac{4\alpha^2}{\pi}\omega \left(2\frac{\arcsin t/2}{t} - 1\right).
    \label{eq:21}
\end{align}
where we made a variable substitution $k = X\sqrt{2\omega}$,
$k_1 = Y\sqrt{2\omega}$ and $t = e^{-\omega x}$.

Now we continue and compute the term
\begin{align}
  - \xi^2 \sum_{\vec k}\sum_{\vec k_1}
  &e^{-\frac{k^2+k_1^2}{4\omega}}
  \frac{u_k u_{k_1}}{k k_1} \frac{(\vec k_{1}\vec k)^2}{8 \omega^2}
  e^{-2\omega x} \nonumber
  \\
  &= -\frac{\alpha^2}{2\pi^4}
  \frac{(4\pi)^2 \omega}{3} \int_{0}^{\infty} \int_{0}^{\infty} dX dY
  e^{- (X^2+ Y^2)}X^2Y^2t^2 \nonumber
  \\
  &= - \frac{\alpha^2}{6\pi}t^2\omega. \label{eq:22}
\end{align}

Finally, we are now able to compute the integrals over $x$
\begin{align}
  I_1
  &= \alpha \sqrt{\frac{\omega}{\pi}}
    \left(\int_0^{\infty}dx\frac{e^{-x}}{\sqrt{1- e^{-\omega x}}}- 1\right),
    \nonumber
  \\
  &= \alpha \sqrt{\frac{\omega}{\pi}}\left(\sqrt{\pi}
    \frac{\Gamma\left(1 + \frac{1}{\omega}\right)}
        {\Gamma\left(\frac{1}{2}+ \frac{1}{\omega}\right)} - 1\right)
    \label{eq:23}
  \\
  I_2
  &= \frac{4\alpha^2}{\pi}\omega\int_0^{\infty}dx \left[\frac{2\arcsin t/2}{t} - 1\right]
    \nonumber
  \\
  &= \{e^{-\omega x} = t; dx = - \frac{dt}{\omega t}\} \nonumber
  \\
  &= \frac{4\alpha^2}{\pi} \int_0^{1}dt \left[\frac{2\arcsin t/2}{t^2} - \frac{1}{t}\right]
    \nonumber
  \\
  &=\frac{4\alpha^2}{\pi} \left(-\frac{\pi}{3} + 1 + 2\ln 2  - \ln(2 + \sqrt{3})\right), \label{eq:24}
  \\
  I_3
  &= - \frac{\alpha^2}{3\pi}\int_0^{\infty}dx t^2\omega = - \frac{\alpha^2}{12\pi}.
    \label{eq:25}
\end{align}

Then the total energy is
\begin{align}
  E_0 (\alpha)
  &\approx E_0^{(0)} + E_0^{(2)} \nonumber
  \\
  &=-\frac{\alpha^2}{3\pi} - (I_1 + I_2 + I_3) \nonumber
  \\
  &= -\frac{\alpha^2}{3\pi}\left(13 + 24 \ln 2 - 4 \pi - 12\ln(2 + \sqrt{3})  - \frac{1}{4}\right)
    \nonumber
  \\
   &\mspace{30mu}+ \alpha \sqrt{\frac{\omega}{\pi}}
     \left(1 - \sqrt{\pi}
    \frac{\Gamma\left(1 + \frac{1}{\omega}\right)}
        {\Gamma\left(\frac{1}{2}+ \frac{1}{\omega}\right)}\right).
     \label{eq:26}
\end{align}

This expression leads to the following  asymptotical limits
\begin{align}
  E_0 (\alpha)
  &\approx - \alpha + 0.1044 \alpha^2 + \ldots,
  \quad  \alpha \rightarrow 0; \label{eq:27}
  \\
  E_0 (\alpha)
  &\approx - 0.1077\alpha^2 - 0.75 \ldots,
  \quad  \alpha \rightarrow \infty. \label{eq:28}
\end{align}

In Fig.~\ref{fig:1} compares the results of both approaches for the
intermediate coupling constant. One can see that our analytical
formula leads to the all-coupling interpolation for the polaron ground
state energy with relative difference less than 15\% in comparison
with Feynman result (Fig.~\ref{fig:1}). Besides, usage of the OM in
this problem allows one to calculate the corrections by means of some
regular procedure \cite{Feranchuk2015}. While for the path-integral
approach the calculation of the subsequent corrections becomes much
more involved. It is important to stress that usage of the resolvent
when calculating the second order correction (\ref{eq:14}) includes
the whole excitation spectrum when summation over the intermediate
states. Possibly it explains why the only trial function can not be
sufficient for the variational solution of the polaron problem.

\section{Calculation of the effective mass}

We have calculated above the binding energy of the rest polaron. In
order to calculate the polaron effective mass, one should consider
this system with nonzero momentum $\vec P \neq 0$. We suppose to solve
this problem on the basis of the OM and formulate it in the
variational form. It is well known that the exact state vector
$|\psi\rangle$ in the Schr\"{o}dinger representation can be found by
variation of the functional
\begin{align}
  J = \langle\psi|[\hat H - E]|\psi\rangle, \label{eq:29}
\end{align}
with additional normalization condition $\langle\psi| \psi\rangle = 1$.

The exact solution should also satisfy to the condition
\begin{align}
  \langle\psi| \hat{\vec I} |\psi\rangle
  &= \vec P, \label{eq:30}
  \\
  \hat{\vec I}
  &= \hat{\vec p} + \sum_{\vec k}\vec k \hat{c}^{\dag}_{\vec k} \hat{c}_{\vec k},
  \label{eq:31}
\end{align}
where $ \vec P $ is the total momentum of the system and $\hat{\vec I}$ is
the corresponding operator, $\hat{\vec p}$ is the electron momentum
operator. If we introduce 3 Lagrange multipliers $\vec V$ then we can
use the only functional
\begin{align}
  J(\vec P) = \langle\psi|[\hat H - E - \vec V \cdot \hat{\vec I}]|\psi\rangle,
  \label{eq:32}
\end{align}
that leads to the following Schr\"{o}dinger equation
\begin{align}
  J(\vec P)
  &= \langle\psi|[\hat H - E - \vec V \cdot \hat{\vec I}]|\psi\rangle,
  \label{eq:33}
  \\
  (\hat H   - \vec V \cdot \hat{\vec I})|\psi\rangle
  &= E |\psi\rangle.
  \label{eq:34}
\end{align}
In case of the slowly moving polaron, one can use the perturbation
theory over the operator $\vec V \cdot \vec I$ together with the OM series
over the operator $\hat H_1$ from Eq.~(\ref{eq:8}). Then the approximate
solution of the Eq.~(\ref{eq:34}) is defined as
\begin{align}
  |\psi\rangle \approx [1 - (\hat H_0 - E_0)^{-1}
  (\hat H_1 -  \vec V \cdot \hat{\vec I})] |\psi_0\rangle,
  \label{eq:35}
\end{align}
with $H_0, |\psi_0>$ from the
Eqs.~(\ref{eq:12}-\ref{eq:13}). Parameters $\vec V$ should be found
from Eq.~(\ref{eq:30}) with the state vector Eq.~(\ref{eq:35})
\begin{align}
  P_\mu
  &= \langle\psi_0| [1 -  (\hat H_1 - \vec V \cdot \vec I)(\hat H_0 - E_0)^{-1}]
  \nonumber
  \\
  &\times I_{\mu}[1 - (\hat H_0 - E_0)^{-1} (\hat H_1 - \vec V\cdot\hat{\vec I})]   |\psi_0\rangle
    \label{eq:36}
\end{align}
and with the considered accuracy
\begin{align}
  P_{\mu}
  &= 2\langle\psi_0| I_{\mu} (\hat H_0 - E_0)^{-1}
    \vec V\cdot\hat{\vec I}|\psi_0\rangle
  \nonumber
  \\
  &- 2\langle\psi_0|\hat H_1 (\hat H_0 - E_0)^{-1} I_{\mu}
    (\hat H_0 - E_0)^{-1}\vec V\cdot\hat{\vec I}|\psi_0\rangle.
    \label{eq:37}
\end{align}

Taking into account the canonical transformations
Eqs.~(\ref{eq:2}-\ref{eq:3}) of variables, one can find in the OM
zeroth approximation for the effective mass of the polaron $m_p$:
\begin{align}
  P_{\lambda}
  &= 2 \langle\psi_0|\hat{I_{\lambda}} (\hat H_0 - E_0)^{-1} \vec V\cdot \hat{\vec I}
    |\psi_0\rangle, \label{eq:38}
  \\
  I_{\lambda}
  &= i \sqrt{\frac{\omega}{2}}(\hat{a}^{\dag}_{\lambda} - \hat{a}_{\lambda})
    + \sum_{\vec k}k_{\lambda} \left(\frac{1}{2}u_k^2
    + \frac{u_k(\hat{b}_{\vec k} + \hat{b}^{\dag}_{\vec k})}
           {\sqrt{2}} + \hat{b}^{\dag}_{\vec k} \hat{b}_{\vec k}\right),
  \label{eq:39}
  \\
  P_{\lambda}
  &= 2 \langle\psi_0|\left(-i \sqrt{\frac{\omega}{2}}  \hat{a}_{\lambda}
    + \sum_{\vec k}k_{\lambda}  \frac{u_k \hat{b}_{\vec k}}{\sqrt{2}}\right)
    \nonumber
  \\
  &\times\int_0^{\infty} dx\left( i \sqrt{\frac{\omega}{2}}\hat{a}^{\dag}_{\mu} e^{- \omega x}
    + \sum_{\vec k}k_{\mu}  \frac{u_k \hat{b}^{\dag}_{\vec k} }{\sqrt{2}}e^{-x}\right)
    V_{\mu}|\psi_0\rangle.
    \label{eq:40}
\end{align}

Parameters $V_{\lambda}$ define 3 components of the ``polaron'' velocity
and the OM zeroth order approximation for its effective mass leads to
\begin{align}
  P^{(0)}_{\lambda}
  &= V_{\lambda}\left[ 1 + \frac{1}{3} \sum_{\vec k}k^2 u_k^2\right],
  \label{eq:41}
  \\
  m_p^{(0)}
  &= 1 + \frac{16 \alpha^4}{81 \pi^2}. \label{eq:42}
\end{align}

The OM correction to the mass can be calculated by the formula
\begin{align}
  P^{(1)}_{\lambda}
  &= - 2\langle\psi_0|  \hat H_1(\hat H_0 - E_0)^{-1}\hat{I}_{\lambda}
  (\hat H_0 - E_0)^{-1}\vec V\cdot\hat{\vec I} |\psi_0\rangle
  \nonumber
  \\
  &- 2\langle\psi_0|\hat H_1\int_0^{\infty}dy e^{-(\hat H_0 - E_0)y}
    \nonumber
  \\
  &\times \hat{I}_{\mu} \int_0^{\infty}dx e^{ - (\hat H_0 - E_0)x}
    \vec V\cdot\hat{\vec I} |\psi_0\rangle.
    \label{eq:43}
\end{align}

For this we compute
\begin{align}
  &\langle\psi_0| \hat H_1 e^{-(\omega \hat{n}+\sum_{\vec k}\hat{N}_{\vec k})y}
    \nonumber
  \\
  &= \xi\langle\psi_0|\sum_{\vec{k_1}}\frac{e^{\frac{-k_1^2}{4\omega}}}{k_1}
    \Bigg[\left( \frac{\hat{b}_{\vec{k_1}}}{\sqrt{2}}e^{-y}  + u_{\vec k1}\right)
    \left(\exp\left( \frac{ik_{1\nu}\hat{a}_\nu e^{-\omega y}}{\sqrt{2\omega}}\right) -1\right)
    \nonumber
  \\
  & \mspace{105mu}+ u_{\vec{k_1}}\frac{k_{1\nu}k_{1\sigma}}{4\omega}
    (\hat{a}_\nu \hat{a}_\sigma)e^{-2\omega y}\Bigg]. \label{eq:44}
\end{align}
and
\begin{align}
  &\hat{I}_{\mu} e^{-(\omega \hat{n}+\sum_{\vec k}\hat{N}_{\vec k})x}\hat{I}_{\nu}
    |\psi_0\rangle
  \nonumber
  \\
  &=\Bigg[i\sqrt{\frac{\omega}{2}}\hat{a}^{\dag}_{\mu}
    + \sum_{\vec{k_2}} \left(\frac{k_{2\mu}u_{k_2} \hat{b}^{\dag}_{\vec{k_2}}}{\sqrt{2}}
    + k_{2\mu}\hat{b}^{\dag}_{k_2} \hat{b}_{ \vec k_2}  \right)\Bigg]
    \nonumber
  \\
  &\times\Bigg[ i\sqrt{\frac{\omega}{2}}\hat{a}^{\dag}_{\nu}e^{-\omega x}
    +\sum_{\vec{k_2}}
    \frac{k_{2\nu}u_{k_2} \hat{b}^{\dag}_{\vec{k_2}}}{\sqrt{2}} e^{-x} \Bigg]
    |\psi_0\rangle. \label{eq:45}
\end{align}

Non zero matrix elements are the following:
\begin{align}
  &\langle\psi_0|\xi\sum_{\vec{k_1}}\frac{e^{\frac{-k_1^2}{4\omega}}}{k_1}
  \frac{\hat{b}_{\vec{k_1}}}{\sqrt{2}} e^{-y}
  \left( \exp\left( \frac{ik_{1\nu}\hat{a}_\nu e^{-\omega y}}{\sqrt{2\omega}}\right) - 1 \right)
  \nonumber
  \\
  &\times \Bigg(i\sqrt{\frac{\omega}{2}}\hat{a}^{\dag}_{\mu}
    \sum_{\vec{k_2}} \frac{k_{2\nu}u_{k_2} \hat{b}^{\dag}_{\vec{k_2}}}{\sqrt{2}}
    e^{-x} \nonumber
  \\
  &\mspace{90mu}+ \sum_{\vec{k_2}}
    \frac{k_{2\mu}u_{k_2} \hat{b}^{\dag}_{\vec{k_2}}}{\sqrt{2}}
  i\sqrt{\frac{\omega}{2}}\hat{a}^{\dag}_{\nu}e^{-\omega x}\Bigg)|\psi_0\rangle 
  \nonumber
  \\
  &= - \xi\frac{1}{12} \delta_{\mu \nu}\sum_{\vec{k }}
  \frac{e^{\frac{-k^2}{4\omega}}}{k_1} e^{-(\omega + 1)y}
  ( e^{-x } + e^{-\omega x})k^2 u_k, \label{eq:46}
\end{align}
and after integrating one can find
\begin{align}
  P_{\mu}^{(1)}
  &= V_{\mu} 2^{5/2}\frac{\alpha}{8\pi^2}\frac{1}{12} \frac{1}{\omega + 1}
  \left( 1 + \frac{1}{\omega}\right) 4 \pi
  \int_0^{\infty} k^2 e^{\frac{-k^2}{2\omega}}dk
  \nonumber
  \\
  &= V_{\mu}\frac{\alpha}{3\omega \pi} \frac{\sqrt{\pi}}{4\sqrt{2}} (2\omega)^{3/2}
    = V_{\mu}\frac{\alpha}{3 }\sqrt{  \frac{\omega}{\pi}}
    = V_{\mu}\frac{2\alpha^2 }{9\pi }.
    \label{eq:47}
\end{align}
Accordingly, the effective mass equals to
\begin{align}
  m_p \approx 1 + \frac{16 \alpha^4}{81 \pi^2} + \frac{2\alpha^2}{9\pi }.
  \label{eq:48}
\end{align}

\begin{figure}
  \includegraphics[width=\columnwidth]{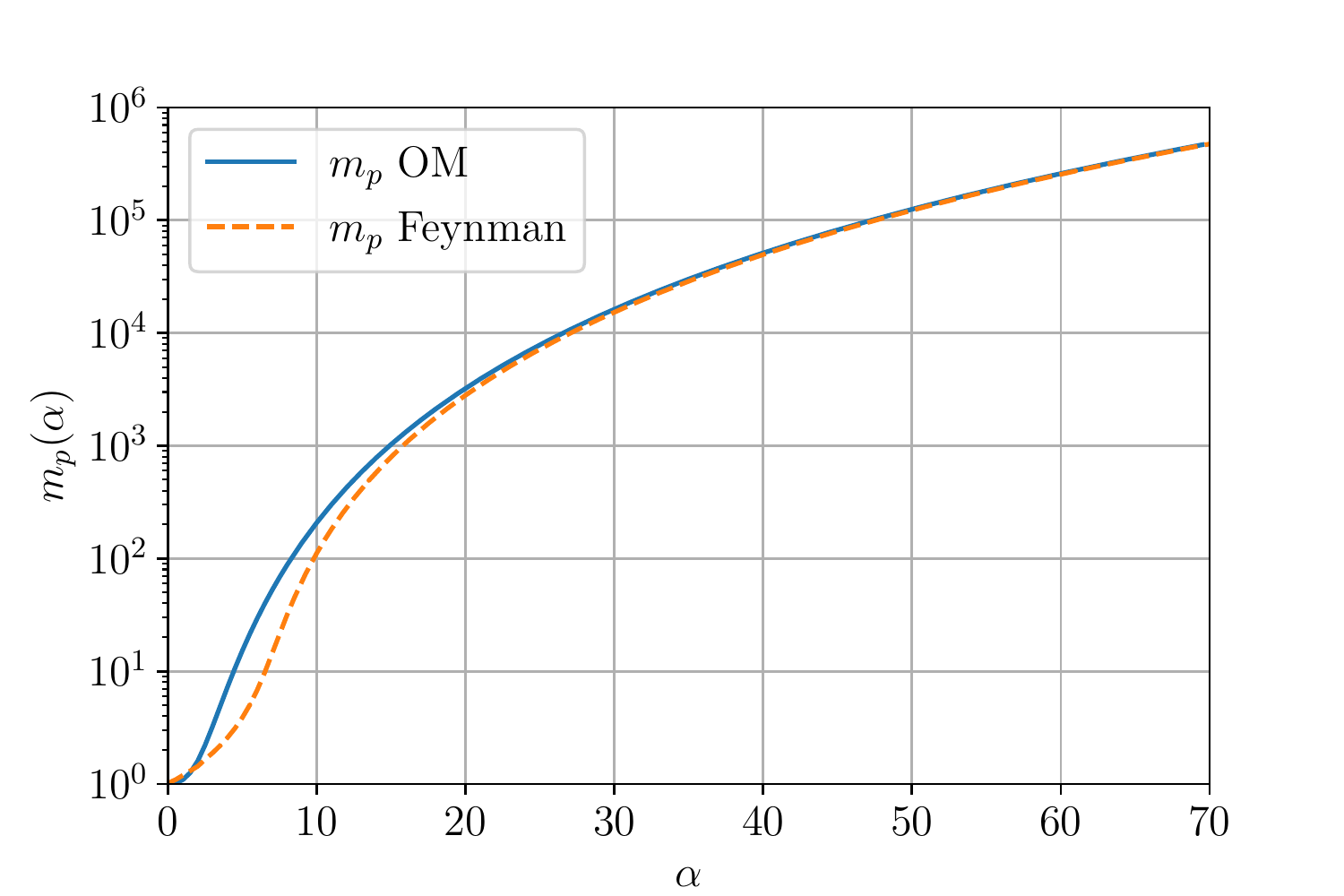}
  \caption{Effective mass as a function of $\alpha$ calculated by
    Feynman and our analytical formula in the logarithmic scale}
  \label{fig:2}
\end{figure}

Fig.~\ref{fig:2} shows that this simple formula leads to all-coupling
approximation for Feynman's result which is connected with rather
complicated variational calculations \cite{feynman1955slow}. Again one
can calculate additional corrections to the effective mass if the
high-order terms on the operator $\hat H_1$ will be taken into account
in the equation (\ref{eq:35}).

\section{Conclusions}

Simple algorithm for calculation of the polaron ground state and its
characterisics in the entire range of the coupling constant is
developed in the frameworks of the Schr\"{o}dinger representation of
the system. The method demands essentially less calculations in
comparison with variational estimation of the functional integrals for
this problem, and leads to the regular procedure for the calculation
of the high-order corrections. It may be useful for other models in
the quantum field theory.

\bibliography{quantum-3}

\end{document}